\documentstyle[12pt]{article}

\title{ Multicomponent nonisothermal nucleation. 3. Numerical results}

\author{V.B.Kurasov}

\date{Victor.Kurasov@pobox.spbu.ru}

\begin{document}

\maketitle

We continue the theory presented in preprints 990958 and 990960 in this
archive. Now all necessary formulas are known and we shall make some calculations
in order to compare our results with the already known theories. All definitions
are the same and no special refereces are given in the text.

\section{Numerical results and conclusions}

To show the numerical effects of the error approach  \cite{Djik} we shall
consider the same situation as it was done in \cite{Djik}.
As far as in has not been declared in \cite{Djik} what normalizing factor
of the equilibrium distribution was used to calculate the stationary rate
of nucleation we have to use the isothermal rate of nucleation published
in \cite{Djik} (see Fig.1 there) as some given data\footnote{The same
qualitative picture will be under the arbitrary normalizing factor.}. The
detailed description of the experimental conditions and data
can be found in \cite{Djik}, \cite{Vis-Strey}.

The condensation of the ethanol (first component) - haxaganol (second
component) is considered. The  nucleation rate logarithm
over the  mean activity $z=\sqrt{\zeta_1^2 + \zeta_2^2}$ is drawn for
the several values of the  activity fraction $q=\zeta_1/ (\zeta_1 + \zeta_2)$.
In Fig.1. the points correspond to the results of Strey and Visanen
\cite{Vis-Strey}.
The solid lines show the isothermal rates of nucleation. Two dashed lines
presents the nonisothermal nucleation rates for different values of the
passive gas (argon)
accommodation coefficient $\alpha_{acc\ g}$. The lower curve corresponds to
$\alpha_{acc\ g} = 0.01$, the upper corresponds to $\alpha_{acc\ g} =
0.1$ (for all activity fractions).

The values of $q$ are written below the series of experimental points
and above the theoretical curves. For small values of $q$ the isothermal
and nonisothermal curves practically coincides, but this occurs only due
to the big slope  of  the  drawn  dependencies.  Moreover  one  can
analytically
show that the difference in $J$ between isothermal and nonisothermal
approaches
is growing with the growth of the nucleation rate and, thus, for small
$q$ this difference is the greatest.

We omit the comparison with the results of Lazaridiz and Drossinos \cite{L-D}
because
their nucleation rates are higher than the classical isothermal  results.
It lies in contradiction with the  principle of stability. It is quite
possible that  Lazaridis
and Drossinos used another input data as the parameters of their theory.

Fig.2 and Fig.3 show the difference between the nucleation rates calculated
by Djikaiev et al. \cite{Djik} and by the formulas presented here. Our
results are dotted lines, the results of Djikaiev et al. are dashed lines,
the nonisothermal rates logarithms are solid lines.
All curves are drawn for $\alpha_{acc\ g} = 0.1$.
The greater the nucleation rate is  the greater is the
manifestation of the thermal effects and the greater is the difference
between the nucleation rate calculated by Djikaiev et al. and our  results.
That's
why we take two situations with the lowest theoretical nucleation rates
which corresponds to $q= 0.980$ (Fig.2) and $q=0.929$ (Fig.3). Certainly, the
difference for $\ln J$ isn't too big, but the correct account of the passive
gas cooling changes  $J$ in several times in comparison with results of
Djikaiev et al.
Our results are closer to the experimental data.

To show the qualitative difference we can assume that $\tau_j$ , $W^+_j$
$n_{\infty \ j}$ $\partial F / \partial \nu_j$
have equal values for all components. Then all components
can not be separated and we have the nonisothermal nucleation for one
component but the passive gas is taken $i_0$ times into account in \cite{Djik}
($i_0$ is the
number of the condensating components). Also we can approximately assume
that the main cooling of the embryo occurs due to the passive gas.
Then taking into account that the renormalization of the stationary rate
is proportional now to the quantity of the passive gas \cite{Kuni-prep}
we can see that
the error in $J$ attains $i_0$ times (two times in the binary condensation).
This error is likely more significant than the difference between the Stauffer
approach \cite{Stauffer} and the steepens   descent method \cite{Reiss}.

All necessary limit transitions of the presented theory (to the one component
theory, to the nonisothermal theory) are observed and give the correct
asymptotes to the already described situations.

To finish our description we can briefly recall the new facts presented
here in comparison with other publications. Certainly, the most advanced
version of the theory  was presented by Djikaiev et al. \cite{Djik},
but even in comparison
with this publication the new features are the following ones:
\begin{itemize}
             \item
The theory is now presented for the multicomponent case.
\item
The shift terms in kinetic equation are obtained. The sense of these terms
is clarified, their negligible role is justified. It is shown that their
negligible role can be shown only in frames of the initial steps of the
Chapman-Enskog procedure. The connection of the vanishing of the shift
terms and the possibility to forget about the lattice structure of the
distribution domain is shown.
\item
The common cooling by the passive instead of the separate cooling is
considered.
This leads to essential numerical difference in the nucleation rate.
\item
The relaxation in the absence of specific parameter required in \cite{Djik}
is based. It allows to consider by the known Chapman-Enskog approach the
situation of the strong thermal effects.
\item
The wrong parameter of decomposition presented in \cite{Djik} is now
corrected.
This clarify the  transition to the isothermal multicomponent theory.

\end{itemize}

The evident weak point of the presented theory is the absence of the surface
tension dependence on the temperature. This phenomena will be taken into
account in the next publication.

\pagebreak

\begin{picture}(400,400)
\put(25,25){\vector(1,0){300}}
\put(25,25){\vector(0,1){300}}
\put(25,330){$\ln J$}
\put(330,25){$z$}
\put(0,0){\line(0,1){350}}
\put(0,0){\line(1,0){350}}
\put(350,0){\line(0,1){350}}
\put(0,350){\line(1,0){350}}
\put(30,35){0.393}
\put(39,49){.}
\put(43,78){.}
\put(44,94){.}
\put(46,97){.}
\put(46,112){.}
\put(47,126){.}
\put(80,40){0.707}
\put(84,50){.}
\put(86,59){.}
\put(88,64){.}
\put(88,67){.}
\put(89,84){.}
\put(93,92){.}
\put(91,98){.}
\put(93,102){.}
\put(93,114){.}
\put(92,121){.}
\put(95,122){.}
\put(96,128){.}
\put(95,137){.}
\put(120,30){0.845}
\put(140,39){.}
\put(142,52){.}
\put(146,73){.}
\put(151,87){.}
\put(153,102){.}
\put(156,110){.}
\put(158,121){.}
\put(166,137){.}
\put(190,40){0.929}
\put(200,50){.}
\put(204,52){.}
\put(210,66){.}
\put(212,80){.}
\put(217,88){.}
\put(223,105){.}
\put(227,115){.}
\put(228,117){.}
\put(233,128){.}
\put(236,135){.}
\put(235,136){.}
\put(240,30){0.980}
\put(247,42){.}
\put(250,44){.}
\put(249,55){.}
\put(254,60){.}
\put(257,66){.}
\put(265,82){.}
\put(268,87){.}
\put(281,102){.}
\put(284,115){.}
\put(285,120){.}
\put(296,127){.}
\put(45,243){.}
\put(45,244){.}
\put(45,245){.}
\put(45,246){.}
\put(45,247){.}
\put(45,249){.}
\put(45,250){.}
\put(45,251){.}
\put(46,252){.}
\put(46,253){.}
\put(46,254){.}
\put(46,255){.}
\put(46,256){.}
\put(46,257){.}
\put(46,258){.}
\put(46,259){.}
\put(47,260){.}
\put(47,261){.}
\put(47,262){.}
\put(47,263){.}
\put(47,265){.}
\put(47,266){.}
\put(47,267){.}
\put(48,268){.}
\put(48,269){.}
\put(48,270){.}
\put(48,271){.}
\put(48,272){.}
\put(48,273){.}
\put(48,274){.}
\put(48,275){.}
\put(49,276){.}
\put(49,277){.}
\put(49,278){.}
\put(49,279){.}
\put(49,280){.}
\put(49,281){.}
\put(49,282){.}
\put(49,283){.}
\put(50,284){.}
\put(50,285){.}
\put(50,286){.}
\put(50,287){.}
\put(50,288){.}
\put(50,288){.}
\put(50,289){.}
\put(50,290){.}
\put(51,291){.}
\put(51,292){.}
\put(51,293){.}
\put(51,294){.}
\put(51,295){.}
\put(51,296){.}
\put(51,297){.}
\put(52,298){.}
\put(52,299){.}
\put(52,300){.}
\put(52,301){.}
\put(52,302){.}
\put(52,302){.}
\put(52,303){.}
\put(52,304){.}
\put(53,305){.}
\put(53,306){.}
\put(53,307){.}
\put(53,308){.}
\put(53,309){.}
\put(53,310){.}
\put(53,310){.}
\put(53,311){.}
\put(50,320){0.393}
\put(85,212){.}
\put(85,214){.}
\put(85,215){.}
\put(85,217){.}
\put(86,218){.}
\put(86,219){.}
\put(86,221){.}
\put(86,222){.}
\put(87,224){.}
\put(87,225){.}
\put(87,227){.}
\put(87,228){.}
\put(88,229){.}
\put(88,231){.}
\put(88,232){.}
\put(88,234){.}
\put(89,235){.}
\put(89,236){.}
\put(89,238){.}
\put(89,239){.}
\put(90,240){.}
\put(90,242){.}
\put(90,243){.}
\put(91,244){.}
\put(91,245){.}
\put(91,247){.}
\put(91,248){.}
\put(92,249){.}
\put(92,250){.}
\put(92,251){.}
\put(92,253){.}
\put(93,254){.}
\put(93,255){.}
\put(93,256){.}
\put(93,257){.}
\put(94,258){.}
\put(94,259){.}
\put(94,261){.}
\put(94,262){.}
\put(95,263){.}
\put(95,264){.}
\put(95,265){.}
\put(95,266){.}
\put(96,267){.}
\put(96,268){.}
\put(96,269){.}
\put(96,270){.}
\put(97,271){.}
\put(97,272){.}
\put(97,273){.}
\put(97,274){.}
\put(98,275){.}
\put(98,276){.}
\put(98,277){.}
\put(98,278){.}
\put(99,278){.}
\put(99,279){.}
\put(99,280){.}
\put(100,281){.}
\put(100,282){.}
\put(100,283){.}
\put(100,284){.}
\put(101,284){.}
\put(90,290){0.707}
\put(133,174){.}
\put(133,175){.}
\put(134,176){.}
\put(134,177){.}
\put(135,179){.}
\put(135,180){.}
\put(135,181){.}
\put(136,182){.}
\put(136,183){.}
\put(136,184){.}
\put(137,186){.}
\put(137,187){.}
\put(138,188){.}
\put(138,189){.}
\put(138,190){.}
\put(139,191){.}
\put(139,192){.}
\put(140,193){.}
\put(140,195){.}
\put(140,196){.}
\put(141,197){.}
\put(141,198){.}
\put(141,199){.}
\put(142,200){.}
\put(142,201){.}
\put(143,202){.}
\put(143,203){.}
\put(143,205){.}
\put(144,206){.}
\put(144,207){.}
\put(145,208){.}
\put(145,209){.}
\put(145,210){.}
\put(146,211){.}
\put(146,212){.}
\put(147,213){.}
\put(147,214){.}
\put(147,215){.}
\put(148,216){.}
\put(148,217){.}
\put(148,218){.}
\put(149,219){.}
\put(149,220){.}
\put(150,221){.}
\put(150,222){.}
\put(150,223){.}
\put(151,224){.}
\put(151,226){.}
\put(152,227){.}
\put(152,228){.}
\put(152,229){.}
\put(153,230){.}
\put(153,231){.}
\put(153,232){.}
\put(154,233){.}
\put(154,233){.}
\put(155,234){.}
\put(155,235){.}
\put(155,236){.}
\put(156,237){.}
\put(156,238){.}
\put(157,239){.}
\put(157,240){.}
\put(157,241){.}
\put(158,242){.}
\put(158,243){.}
\put(158,244){.}
\put(159,245){.}
\put(159,246){.}
\put(160,247){.}
\put(160,248){.}
\put(160,249){.}
\put(161,250){.}
\put(161,251){.}
\put(162,252){.}
\put(162,252){.}
\put(162,253){.}
\put(163,254){.}
\put(163,255){.}
\put(164,256){.}
\put(164,257){.}
\put(164,258){.}
\put(165,259){.}
\put(165,260){.}
\put(150,265){0.845}
\put(197,159){.}
\put(198,160){.}
\put(199,161){.}
\put(199,163){.}
\put(200,164){.}
\put(200,165){.}
\put(201,166){.}
\put(201,167){.}
\put(202,168){.}
\put(202,169){.}
\put(203,170){.}
\put(203,171){.}
\put(204,172){.}
\put(204,173){.}
\put(205,174){.}
\put(205,175){.}
\put(206,176){.}
\put(206,177){.}
\put(207,178){.}
\put(207,179){.}
\put(208,180){.}
\put(208,181){.}
\put(209,182){.}
\put(209,183){.}
\put(210,184){.}
\put(210,185){.}
\put(211,186){.}
\put(211,187){.}
\put(212,188){.}
\put(212,189){.}
\put(213,190){.}
\put(213,191){.}
\put(214,192){.}
\put(214,193){.}
\put(215,194){.}
\put(216,195){.}
\put(216,196){.}
\put(217,197){.}
\put(217,198){.}
\put(218,199){.}
\put(218,200){.}
\put(219,201){.}
\put(219,202){.}
\put(220,203){.}
\put(220,204){.}
\put(221,204){.}
\put(221,205){.}
\put(222,206){.}
\put(222,207){.}
\put(223,208){.}
\put(223,209){.}
\put(224,210){.}
\put(224,211){.}
\put(225,212){.}
\put(225,213){.}
\put(226,214){.}
\put(226,215){.}
\put(227,216){.}
\put(227,217){.}
\put(228,218){.}
\put(228,219){.}
\put(229,220){.}
\put(229,221){.}
\put(230,222){.}
\put(230,222){.}
\put(231,223){.}
\put(231,224){.}
\put(232,225){.}
\put(233,226){.}
\put(233,227){.}
\put(234,228){.}
\put(234,229){.}
\put(235,230){.}
\put(235,231){.}
\put(236,232){.}
\put(230,240){0.929}
\put(250,135){.}
\put(250,136){.}
\put(251,137){.}
\put(252,138){.}
\put(252,139){.}
\put(253,140){.}
\put(254,141){.}
\put(254,142){.}
\put(255,143){.}
\put(256,144){.}
\put(256,145){.}
\put(257,146){.}
\put(257,147){.}
\put(258,148){.}
\put(259,149){.}
\put(259,150){.}
\put(260,151){.}
\put(261,152){.}
\put(261,153){.}
\put(262,154){.}
\put(263,155){.}
\put(263,156){.}
\put(264,157){.}
\put(265,158){.}
\put(265,159){.}
\put(266,160){.}
\put(266,161){.}
\put(267,162){.}
\put(268,163){.}
\put(268,164){.}
\put(269,165){.}
\put(270,166){.}
\put(270,167){.}
\put(271,168){.}
\put(272,169){.}
\put(272,170){.}
\put(273,171){.}
\put(274,172){.}
\put(274,173){.}
\put(275,174){.}
\put(276,174){.}
\put(276,175){.}
\put(277,176){.}
\put(277,177){.}
\put(278,178){.}
\put(279,179){.}
\put(279,180){.}
\put(280,181){.}
\put(281,182){.}
\put(281,183){.}
\put(282,184){.}
\put(283,185){.}
\put(283,186){.}
\put(284,186){.}
\put(285,187){.}
\put(285,188){.}
\put(286,189){.}
\put(286,190){.}
\put(287,191){.}
\put(288,192){.}
\put(288,193){.}
\put(289,194){.}
\put(290,194){.}
\put(290,195){.}
\put(291,196){.}
\put(292,197){.}
\put(292,198){.}
\put(293,199){.}
\put(294,200){.}
\put(290,210){0.980}
\put(46,243){.}
\put(46,244){.}
\put(46,245){.}
\put(46,246){.}
\put(46,247){.}
\put(46,249){.}
\put(46,250){.}
\put(47,251){.}
\put(47,252){.}
\put(47,253){.}
\put(47,254){.}
\put(47,255){.}
\put(47,256){.}
\put(47,257){.}
\put(47,258){.}
\put(48,259){.}
\put(48,260){.}
\put(48,261){.}
\put(48,262){.}
\put(48,263){.}
\put(48,265){.}
\put(48,266){.}
\put(52,291){.}
\put(52,292){.}
\put(52,293){.}
\put(52,294){.}
\put(52,295){.}
\put(52,296){.}
\put(52,297){.}
\put(53,298){.}
\put(53,299){.}
\put(53,300){.}
\put(53,301){.}
\put(53,302){.}
\put(53,302){.}
\put(53,303){.}
\put(53,304){.}
\put(54,305){.}
\put(54,306){.}
\put(54,307){.}
\put(54,308){.}
\put(54,309){.}
\put(54,310){.}
\put(54,310){.}
\put(55,311){.}
\put(86,212){.}
\put(86,214){.}
\put(86,215){.}
\put(86,217){.}
\put(87,218){.}
\put(87,219){.}
\put(90,238){.}
\put(91,239){.}
\put(91,240){.}
\put(91,242){.}
\put(91,243){.}
\put(92,244){.}
\put(92,245){.}
\put(92,247){.}
\put(92,248){.}
\put(93,249){.}
\put(93,250){.}
\put(93,251){.}
\put(93,253){.}
\put(97,267){.}
\put(97,268){.}
\put(97,269){.}
\put(97,270){.}
\put(98,271){.}
\put(98,272){.}
\put(98,273){.}
\put(99,274){.}
\put(99,275){.}
\put(99,276){.}
\put(99,277){.}
\put(100,278){.}
\put(100,278){.}
\put(139,179){.}
\put(139,180){.}
\put(140,181){.}
\put(140,182){.}
\put(140,183){.}
\put(141,184){.}
\put(141,186){.}
\put(141,187){.}
\put(145,197){.}
\put(145,198){.}
\put(146,199){.}
\put(146,200){.}
\put(147,201){.}
\put(147,202){.}
\put(147,203){.}
\put(148,205){.}
\put(148,206){.}
\put(152,215){.}
\put(152,216){.}
\put(152,217){.}
\put(153,218){.}
\put(153,219){.}
\put(153,220){.}
\put(154,221){.}
\put(154,222){.}
\put(158,233){.}
\put(158,233){.}
\put(159,234){.}
\put(159,235){.}
\put(160,236){.}
\put(160,237){.}
\put(160,238){.}
\put(161,239){.}
\put(164,248){.}
\put(165,249){.}
\put(165,250){.}
\put(165,251){.}
\put(166,252){.}
\put(166,252){.}
\put(167,253){.}
\put(167,254){.}
\put(167,255){.}
\put(209,163){.}
\put(209,164){.}
\put(210,165){.}
\put(210,166){.}
\put(211,167){.}
\put(211,168){.}
\put(215,175){.}
\put(216,176){.}
\put(216,177){.}
\put(217,178){.}
\put(217,179){.}
\put(218,180){.}
\put(222,188){.}
\put(222,189){.}
\put(223,190){.}
\put(223,191){.}
\put(224,192){.}
\put(224,193){.}
\put(228,200){.}
\put(228,201){.}
\put(229,202){.}
\put(229,203){.}
\put(230,204){.}
\put(230,204){.}
\put(235,212){.}
\put(235,213){.}
\put(236,214){.}
\put(236,215){.}
\put(237,216){.}
\put(237,217){.}
\put(241,223){.}
\put(241,224){.}
\put(242,225){.}
\put(242,226){.}
\put(243,227){.}
\put(243,228){.}
\put(270,136){.}
\put(271,137){.}
\put(271,138){.}
\put(272,139){.}
\put(273,140){.}
\put(277,146){.}
\put(277,147){.}
\put(278,148){.}
\put(279,149){.}
\put(279,150){.}
\put(283,156){.}
\put(284,157){.}
\put(284,158){.}
\put(285,159){.}
\put(286,160){.}
\put(289,166){.}
\put(290,167){.}
\put(291,168){.}
\put(291,169){.}
\put(292,170){.}
\put(296,175){.}
\put(297,176){.}
\put(297,177){.}
\put(298,178){.}
\put(298,179){.}
\put(302,185){.}
\put(303,186){.}
\put(304,186){.}
\put(304,187){.}
\put(305,188){.}
\put(309,194){.}
\put(309,194){.}
\put(310,195){.}
\put(311,196){.}
\put(311,197){.}
\put(45,243){.}
\put(45,244){.}
\put(46,245){.}
\put(46,246){.}
\put(46,247){.}
\put(46,249){.}
\put(46,250){.}
\put(46,251){.}
\put(46,252){.}
\put(46,253){.}
\put(47,254){.}
\put(47,255){.}
\put(47,256){.}
\put(47,257){.}
\put(47,258){.}
\put(47,259){.}
\put(47,260){.}
\put(47,261){.}
\put(48,262){.}
\put(48,263){.}
\put(48,265){.}
\put(48,266){.}
\put(51,291){.}
\put(51,292){.}
\put(52,293){.}
\put(52,294){.}
\put(52,295){.}
\put(52,296){.}
\put(52,297){.}
\put(52,298){.}
\put(52,299){.}
\put(53,300){.}
\put(53,301){.}
\put(53,302){.}
\put(53,302){.}
\put(53,303){.}
\put(53,304){.}
\put(53,305){.}
\put(53,306){.}
\put(54,307){.}
\put(54,308){.}
\put(54,309){.}
\put(54,310){.}
\put(54,310){.}
\put(54,311){.}
\put(85,212){.}
\put(86,214){.}
\put(86,215){.}
\put(86,217){.}
\put(86,218){.}
\put(87,219){.}
\put(90,238){.}
\put(90,239){.}
\put(90,240){.}
\put(91,242){.}
\put(91,243){.}
\put(91,244){.}
\put(92,245){.}
\put(92,247){.}
\put(92,248){.}
\put(92,249){.}
\put(93,250){.}
\put(93,251){.}
\put(93,253){.}
\put(96,267){.}
\put(97,268){.}
\put(97,269){.}
\put(97,270){.}
\put(97,271){.}
\put(98,272){.}
\put(98,273){.}
\put(98,274){.}
\put(98,275){.}
\put(99,276){.}
\put(99,277){.}
\put(99,278){.}
\put(99,278){.}
\put(138,179){.}
\put(138,180){.}
\put(138,181){.}
\put(139,182){.}
\put(139,183){.}
\put(139,184){.}
\put(140,186){.}
\put(140,187){.}
\put(144,197){.}
\put(144,198){.}
\put(144,199){.}
\put(145,200){.}
\put(145,201){.}
\put(146,202){.}
\put(146,203){.}
\put(146,205){.}
\put(147,206){.}
\put(150,215){.}
\put(151,216){.}
\put(151,217){.}
\put(151,218){.}
\put(152,219){.}
\put(152,220){.}
\put(153,221){.}
\put(153,222){.}
\put(157,233){.}
\put(157,233){.}
\put(158,234){.}
\put(158,235){.}
\put(158,236){.}
\put(159,237){.}
\put(159,238){.}
\put(160,239){.}
\put(163,248){.}
\put(163,249){.}
\put(164,250){.}
\put(164,251){.}
\put(165,252){.}
\put(165,252){.}
\put(165,253){.}
\put(166,254){.}
\put(166,255){.}
\put(206,163){.}
\put(207,164){.}
\put(207,165){.}
\put(208,166){.}
\put(208,167){.}
\put(209,168){.}
\put(212,175){.}
\put(213,176){.}
\put(213,177){.}
\put(214,178){.}
\put(214,179){.}
\put(215,180){.}
\put(219,188){.}
\put(220,189){.}
\put(220,190){.}
\put(221,191){.}
\put(221,192){.}
\put(222,193){.}
\put(225,200){.}
\put(226,201){.}
\put(226,202){.}
\put(227,203){.}
\put(227,204){.}
\put(228,204){.}
\put(232,212){.}
\put(233,213){.}
\put(233,214){.}
\put(234,215){.}
\put(234,216){.}
\put(235,217){.}
\put(238,223){.}
\put(239,224){.}
\put(239,225){.}
\put(240,226){.}
\put(240,227){.}
\put(241,228){.}
\put(264,136){.}
\put(265,137){.}
\put(266,138){.}
\put(266,139){.}
\put(267,140){.}
\put(271,146){.}
\put(271,147){.}
\put(272,148){.}
\put(273,149){.}
\put(273,150){.}
\put(277,156){.}
\put(278,157){.}
\put(279,158){.}
\put(279,159){.}
\put(280,160){.}
\put(284,166){.}
\put(284,167){.}
\put(285,168){.}
\put(286,169){.}
\put(286,170){.}
\put(290,175){.}
\put(291,176){.}
\put(291,177){.}
\put(292,178){.}
\put(293,179){.}
\put(297,185){.}
\put(297,186){.}
\put(298,186){.}
\put(298,187){.}
\put(299,188){.}
\put(303,194){.}
\put(304,194){.}
\put(304,195){.}
\put(305,196){.}
\put(306,197){.}
\put(53,20){\line(0,1){5}}
\put(53,10){4}
\put(126,20){\line(0,1){5}}
\put(126,10){8}
\put(199,20){\line(0,1){5}}
\put(199,10){12}
\put(272,20){\line(0,1){5}}
\put(272,10){16}
\put(20,54){\line(1,0){5}}
\put(10,54){15}
\put(20,129){\line(1,0){5}}
\put(10,129){20}
\put(20,204){\line(1,0){5}}
\put(10,204){25}
\put(20,279){\line(1,0){5}}
\put(10,279){30}
\end{picture}

{\center Figure 1. }

\pagebreak

\begin{picture}(400,400)
\put(25,25){\vector(1,0){300}}
\put(25,25){\vector(0,1){300}}
\put(25,330){$\ln J$}
\put(330,25){$z$}
\put(0,0){\line(0,1){350}}
\put(0,0){\line(1,0){350}}
\put(350,0){\line(0,1){350}}
\put(0,350){\line(1,0){350}}
\put(40,40){.}
\put(40,41){.}
\put(41,42){.}
\put(42,43){.}
\put(42,44){.}
\put(43,45){.}
\put(44,46){.}
\put(44,47){.}
\put(45,48){.}
\put(46,49){.}
\put(47,50){.}
\put(47,51){.}
\put(48,52){.}
\put(49,53){.}
\put(49,54){.}
\put(50,55){.}
\put(51,56){.}
\put(51,57){.}
\put(52,58){.}
\put(53,59){.}
\put(53,60){.}
\put(54,61){.}
\put(55,62){.}
\put(55,62){.}
\put(55,63){.}
\put(56,64){.}
\put(57,65){.}
\put(57,66){.}
\put(58,67){.}
\put(59,68){.}
\put(60,69){.}
\put(60,70){.}
\put(61,71){.}
\put(62,72){.}
\put(62,73){.}
\put(63,74){.}
\put(64,75){.}
\put(64,76){.}
\put(65,77){.}
\put(66,78){.}
\put(67,79){.}
\put(67,80){.}
\put(68,81){.}
\put(69,82){.}
\put(69,83){.}
\put(70,84){.}
\put(71,85){.}
\put(71,85){.}
\put(71,86){.}
\put(72,87){.}
\put(73,88){.}
\put(74,89){.}
\put(74,90){.}
\put(75,91){.}
\put(76,92){.}
\put(77,93){.}
\put(77,94){.}
\put(78,95){.}
\put(79,96){.}
\put(79,97){.}
\put(80,98){.}
\put(81,99){.}
\put(82,100){.}
\put(82,101){.}
\put(83,102){.}
\put(84,103){.}
\put(85,104){.}
\put(85,105){.}
\put(86,106){.}
\put(87,107){.}
\put(87,107){.}
\put(87,108){.}
\put(88,109){.}
\put(89,110){.}
\put(90,111){.}
\put(90,112){.}
\put(91,113){.}
\put(92,114){.}
\put(93,115){.}
\put(93,116){.}
\put(94,117){.}
\put(95,118){.}
\put(95,119){.}
\put(96,120){.}
\put(97,121){.}
\put(98,122){.}
\put(98,123){.}
\put(99,124){.}
\put(100,125){.}
\put(101,126){.}
\put(101,127){.}
\put(102,128){.}
\put(103,129){.}
\put(103,129){.}
\put(103,130){.}
\put(104,131){.}
\put(105,132){.}
\put(106,133){.}
\put(106,134){.}
\put(107,135){.}
\put(108,136){.}
\put(108,137){.}
\put(109,138){.}
\put(110,139){.}
\put(111,140){.}
\put(111,141){.}
\put(112,142){.}
\put(113,143){.}
\put(113,144){.}
\put(114,145){.}
\put(115,146){.}
\put(116,147){.}
\put(116,148){.}
\put(117,149){.}
\put(118,150){.}
\put(118,150){.}
\put(119,151){.}
\put(119,152){.}
\put(120,153){.}
\put(121,154){.}
\put(122,155){.}
\put(122,156){.}
\put(123,157){.}
\put(124,158){.}
\put(125,159){.}
\put(125,160){.}
\put(126,161){.}
\put(127,162){.}
\put(128,163){.}
\put(128,164){.}
\put(129,165){.}
\put(130,166){.}
\put(131,167){.}
\put(131,168){.}
\put(132,169){.}
\put(133,170){.}
\put(134,171){.}
\put(134,171){.}
\put(135,172){.}
\put(135,173){.}
\put(136,174){.}
\put(137,175){.}
\put(138,176){.}
\put(138,177){.}
\put(139,178){.}
\put(140,179){.}
\put(141,180){.}
\put(141,181){.}
\put(142,182){.}
\put(143,183){.}
\put(144,184){.}
\put(144,185){.}
\put(145,186){.}
\put(146,187){.}
\put(147,188){.}
\put(147,189){.}
\put(148,190){.}
\put(149,191){.}
\put(150,192){.}
\put(150,192){.}
\put(150,193){.}
\put(151,194){.}
\put(152,195){.}
\put(153,196){.}
\put(153,197){.}
\put(154,198){.}
\put(155,199){.}
\put(155,200){.}
\put(156,201){.}
\put(157,202){.}
\put(158,203){.}
\put(158,204){.}
\put(159,205){.}
\put(160,206){.}
\put(160,207){.}
\put(161,208){.}
\put(162,209){.}
\put(163,210){.}
\put(163,211){.}
\put(164,212){.}
\put(165,213){.}
\put(165,213){.}
\put(166,214){.}
\put(166,215){.}
\put(167,216){.}
\put(168,217){.}
\put(169,218){.}
\put(170,219){.}
\put(170,220){.}
\put(171,221){.}
\put(172,222){.}
\put(173,223){.}
\put(174,224){.}
\put(174,225){.}
\put(175,226){.}
\put(176,227){.}
\put(177,228){.}
\put(178,229){.}
\put(178,230){.}
\put(179,231){.}
\put(180,232){.}
\put(181,233){.}
\put(181,233){.}
\put(182,234){.}
\put(182,235){.}
\put(183,236){.}
\put(184,237){.}
\put(185,238){.}
\put(185,239){.}
\put(186,240){.}
\put(187,241){.}
\put(188,242){.}
\put(188,243){.}
\put(189,244){.}
\put(190,245){.}
\put(191,246){.}
\put(191,247){.}
\put(192,248){.}
\put(193,249){.}
\put(194,250){.}
\put(194,251){.}
\put(195,252){.}
\put(196,253){.}
\put(197,254){.}
\put(197,254){.}
\put(198,255){.}
\put(199,256){.}
\put(199,257){.}
\put(200,258){.}
\put(201,259){.}
\put(202,260){.}
\put(203,261){.}
\put(204,262){.}
\put(204,263){.}
\put(205,264){.}
\put(206,265){.}
\put(207,266){.}
\put(208,267){.}
\put(209,268){.}
\put(209,269){.}
\put(210,270){.}
\put(211,271){.}
\put(212,272){.}
\put(213,273){.}
\put(213,273){.}
\put(214,274){.}
\put(214,275){.}
\put(215,276){.}
\put(216,277){.}
\put(217,278){.}
\put(217,279){.}
\put(218,280){.}
\put(219,281){.}
\put(220,282){.}
\put(220,283){.}
\put(221,284){.}
\put(222,285){.}
\put(223,286){.}
\put(223,287){.}
\put(224,288){.}
\put(225,289){.}
\put(226,290){.}
\put(226,291){.}
\put(227,292){.}
\put(228,293){.}
\put(228,293){.}
\put(229,294){.}
\put(230,295){.}
\put(230,296){.}
\put(231,297){.}
\put(232,298){.}
\put(233,299){.}
\put(234,300){.}
\put(235,301){.}
\put(235,302){.}
\put(236,303){.}
\put(237,304){.}
\put(238,305){.}
\put(239,306){.}
\put(240,307){.}
\put(240,308){.}
\put(241,309){.}
\put(242,310){.}
\put(243,311){.}
\put(244,312){.}
\put(110,43){.}
\put(112,46){.}
\put(114,49){.}
\put(116,52){.}
\put(119,55){.}
\put(121,58){.}
\put(123,61){.}
\put(126,65){.}
\put(128,68){.}
\put(130,71){.}
\put(131,74){.}
\put(133,77){.}
\put(135,80){.}
\put(137,83){.}
\put(141,88){.}
\put(143,91){.}
\put(145,94){.}
\put(147,97){.}
\put(150,100){.}
\put(152,103){.}
\put(154,106){.}
\put(157,110){.}
\put(159,113){.}
\put(161,116){.}
\put(163,119){.}
\put(166,122){.}
\put(168,125){.}
\put(170,128){.}
\put(173,132){.}
\put(175,135){.}
\put(178,138){.}
\put(180,141){.}
\put(182,144){.}
\put(184,147){.}
\put(187,150){.}
\put(189,153){.}
\put(191,156){.}
\put(193,159){.}
\put(195,162){.}
\put(197,165){.}
\put(200,168){.}
\put(202,171){.}
\put(204,174){.}
\put(206,177){.}
\put(209,180){.}
\put(211,183){.}
\put(213,186){.}
\put(215,189){.}
\put(218,192){.}
\put(220,195){.}
\put(222,198){.}
\put(225,201){.}
\put(227,204){.}
\put(229,207){.}
\put(231,210){.}
\put(234,213){.}
\put(236,216){.}
\put(238,219){.}
\put(241,222){.}
\put(243,225){.}
\put(245,228){.}
\put(247,231){.}
\put(251,236){.}
\put(253,239){.}
\put(256,242){.}
\put(258,245){.}
\put(260,248){.}
\put(262,251){.}
\put(265,254){.}
\put(267,257){.}
\put(270,260){.}
\put(272,263){.}
\put(275,266){.}
\put(277,269){.}
\put(280,272){.}
\put(283,276){.}
\put(286,279){.}
\put(288,282){.}
\put(290,285){.}
\put(293,288){.}
\put(295,291){.}
\put(299,296){.}
\put(302,299){.}
\put(304,302){.}
\put(306,305){.}
\put(309,308){.}
\put(311,311){.}
\put(100,42){.}
\put(101,43){.}
\put(106,50){.}
\put(107,51){.}
\put(108,53){.}
\put(109,54){.}
\put(114,61){.}
\put(115,62){.}
\put(116,64){.}
\put(117,65){.}
\put(122,72){.}
\put(123,74){.}
\put(124,75){.}
\put(125,77){.}
\put(130,84){.}
\put(131,85){.}
\put(132,87){.}
\put(133,88){.}
\put(138,95){.}
\put(139,96){.}
\put(140,98){.}
\put(141,99){.}
\put(146,106){.}
\put(147,107){.}
\put(148,109){.}
\put(149,110){.}
\put(154,117){.}
\put(155,119){.}
\put(156,120){.}
\put(157,122){.}
\put(161,128){.}
\put(162,129){.}
\put(163,130){.}
\put(164,132){.}
\put(169,138){.}
\put(170,140){.}
\put(171,141){.}
\put(172,142){.}
\put(177,149){.}
\put(178,150){.}
\put(179,151){.}
\put(180,153){.}
\put(185,159){.}
\put(186,161){.}
\put(187,162){.}
\put(188,163){.}
\put(193,170){.}
\put(194,171){.}
\put(195,173){.}
\put(196,174){.}
\put(201,181){.}
\put(202,182){.}
\put(203,184){.}
\put(204,185){.}
\put(208,191){.}
\put(209,192){.}
\put(210,193){.}
\put(211,195){.}
\put(216,201){.}
\put(217,203){.}
\put(218,204){.}
\put(219,205){.}
\put(224,212){.}
\put(225,213){.}
\put(226,214){.}
\put(227,216){.}
\put(232,222){.}
\put(233,223){.}
\put(234,224){.}
\put(235,226){.}
\put(240,232){.}
\put(241,233){.}
\put(242,234){.}
\put(243,236){.}
\put(248,242){.}
\put(249,244){.}
\put(250,245){.}
\put(251,246){.}
\put(256,253){.}
\put(257,254){.}
\put(258,255){.}
\put(259,257){.}
\put(264,263){.}
\put(265,264){.}
\put(266,266){.}
\put(267,267){.}
\put(271,272){.}
\put(272,273){.}
\put(273,274){.}
\put(274,276){.}
\put(279,282){.}
\put(280,283){.}
\put(281,284){.}
\put(282,286){.}
\put(287,292){.}
\put(288,293){.}
\put(289,294){.}
\put(290,295){.}
\put(295,301){.}
\put(296,303){.}
\put(297,304){.}
\put(298,305){.}
\put(303,311){.}
\put(304,312){.}
\put(44,20){\line(0,1){5}}
\put(44,10){15}
\put(133,20){\line(0,1){5}}
\put(133,10){16}
\put(222,20){\line(0,1){5}}
\put(222,10){17}
\put(311,20){\line(0,1){5}}
\put(311,10){18}
\put(20,61){\line(1,0){5}}
\put(10,61){21}
\put(20,127){\line(1,0){5}}
\put(10,127){22}
\put(20,193){\line(1,0){5}}
\put(10,193){23}
\put(20,259){\line(1,0){5}}
\put(10,259){24}
\end{picture}

{\center Figure 2. }

\pagebreak

\begin{picture}(400,400)
\put(25,25){\vector(1,0){300}}
\put(25,25){\vector(0,1){300}}
\put(25,330){$\ln J$}
\put(330,25){$z$}
\put(0,0){\line(0,1){350}}
\put(0,0){\line(1,0){350}}
\put(350,0){\line(0,1){350}}
\put(0,350){\line(1,0){350}}
\put(40,40){.}
\put(41,41){.}
\put(41,42){.}
\put(42,43){.}
\put(43,44){.}
\put(44,45){.}
\put(45,46){.}
\put(45,47){.}
\put(46,48){.}
\put(47,49){.}
\put(48,50){.}
\put(49,51){.}
\put(49,52){.}
\put(50,53){.}
\put(51,54){.}
\put(52,55){.}
\put(53,56){.}
\put(53,57){.}
\put(54,58){.}
\put(55,59){.}
\put(56,60){.}
\put(56,60){.}
\put(57,61){.}
\put(57,62){.}
\put(58,63){.}
\put(59,64){.}
\put(60,65){.}
\put(60,66){.}
\put(61,67){.}
\put(62,68){.}
\put(63,69){.}
\put(63,70){.}
\put(64,71){.}
\put(65,72){.}
\put(66,73){.}
\put(66,74){.}
\put(67,75){.}
\put(68,76){.}
\put(69,77){.}
\put(69,78){.}
\put(70,79){.}
\put(71,80){.}
\put(72,81){.}
\put(72,81){.}
\put(73,82){.}
\put(73,83){.}
\put(74,84){.}
\put(75,85){.}
\put(76,86){.}
\put(77,87){.}
\put(77,88){.}
\put(78,89){.}
\put(79,90){.}
\put(80,91){.}
\put(81,92){.}
\put(81,93){.}
\put(82,94){.}
\put(83,95){.}
\put(84,96){.}
\put(85,97){.}
\put(85,98){.}
\put(86,99){.}
\put(87,100){.}
\put(88,101){.}
\put(88,101){.}
\put(89,102){.}
\put(90,103){.}
\put(90,104){.}
\put(91,105){.}
\put(92,106){.}
\put(93,107){.}
\put(94,108){.}
\put(95,109){.}
\put(96,110){.}
\put(96,111){.}
\put(97,112){.}
\put(98,113){.}
\put(99,114){.}
\put(100,115){.}
\put(101,116){.}
\put(101,117){.}
\put(102,118){.}
\put(103,119){.}
\put(104,120){.}
\put(105,121){.}
\put(105,121){.}
\put(106,122){.}
\put(106,123){.}
\put(107,124){.}
\put(108,125){.}
\put(109,126){.}
\put(110,127){.}
\put(110,128){.}
\put(111,129){.}
\put(112,130){.}
\put(113,131){.}
\put(114,132){.}
\put(114,133){.}
\put(115,134){.}
\put(116,135){.}
\put(117,136){.}
\put(118,137){.}
\put(118,138){.}
\put(119,139){.}
\put(120,140){.}
\put(121,141){.}
\put(121,141){.}
\put(122,142){.}
\put(122,143){.}
\put(123,144){.}
\put(124,145){.}
\put(125,146){.}
\put(126,147){.}
\put(126,148){.}
\put(127,149){.}
\put(128,150){.}
\put(129,151){.}
\put(130,152){.}
\put(130,153){.}
\put(131,154){.}
\put(132,155){.}
\put(133,156){.}
\put(134,157){.}
\put(134,158){.}
\put(135,159){.}
\put(136,160){.}
\put(137,161){.}
\put(137,161){.}
\put(138,162){.}
\put(139,163){.}
\put(139,164){.}
\put(140,165){.}
\put(141,166){.}
\put(142,167){.}
\put(143,168){.}
\put(144,169){.}
\put(144,170){.}
\put(145,171){.}
\put(146,172){.}
\put(147,173){.}
\put(148,174){.}
\put(149,175){.}
\put(149,176){.}
\put(150,177){.}
\put(151,178){.}
\put(152,179){.}
\put(153,180){.}
\put(153,180){.}
\put(154,181){.}
\put(154,182){.}
\put(155,183){.}
\put(156,184){.}
\put(157,185){.}
\put(158,186){.}
\put(158,187){.}
\put(159,188){.}
\put(160,189){.}
\put(161,190){.}
\put(162,191){.}
\put(162,192){.}
\put(163,193){.}
\put(164,194){.}
\put(165,195){.}
\put(166,196){.}
\put(166,197){.}
\put(167,198){.}
\put(168,199){.}
\put(169,200){.}
\put(169,200){.}
\put(170,201){.}
\put(171,202){.}
\put(172,203){.}
\put(172,204){.}
\put(173,205){.}
\put(174,206){.}
\put(175,207){.}
\put(176,208){.}
\put(177,209){.}
\put(178,210){.}
\put(179,211){.}
\put(180,212){.}
\put(181,213){.}
\put(181,214){.}
\put(182,215){.}
\put(183,216){.}
\put(184,217){.}
\put(185,218){.}
\put(186,219){.}
\put(186,219){.}
\put(187,220){.}
\put(188,221){.}
\put(188,222){.}
\put(189,223){.}
\put(190,224){.}
\put(191,225){.}
\put(192,226){.}
\put(193,227){.}
\put(193,228){.}
\put(194,229){.}
\put(195,230){.}
\put(196,231){.}
\put(197,232){.}
\put(198,233){.}
\put(198,234){.}
\put(199,235){.}
\put(200,236){.}
\put(201,237){.}
\put(202,238){.}
\put(202,238){.}
\put(203,239){.}
\put(204,240){.}
\put(204,241){.}
\put(205,242){.}
\put(206,243){.}
\put(207,244){.}
\put(208,245){.}
\put(209,246){.}
\put(209,247){.}
\put(210,248){.}
\put(211,249){.}
\put(212,250){.}
\put(213,251){.}
\put(214,252){.}
\put(214,253){.}
\put(215,254){.}
\put(216,255){.}
\put(217,256){.}
\put(218,257){.}
\put(218,257){.}
\put(219,258){.}
\put(220,259){.}
\put(220,260){.}
\put(221,261){.}
\put(222,262){.}
\put(223,263){.}
\put(224,264){.}
\put(225,265){.}
\put(225,266){.}
\put(226,267){.}
\put(227,268){.}
\put(228,269){.}
\put(229,270){.}
\put(230,271){.}
\put(230,272){.}
\put(231,273){.}
\put(232,274){.}
\put(233,275){.}
\put(234,276){.}
\put(234,276){.}
\put(235,277){.}
\put(236,278){.}
\put(237,279){.}
\put(238,280){.}
\put(239,281){.}
\put(240,282){.}
\put(241,283){.}
\put(242,284){.}
\put(242,285){.}
\put(243,286){.}
\put(244,287){.}
\put(245,288){.}
\put(246,289){.}
\put(247,290){.}
\put(248,291){.}
\put(249,292){.}
\put(250,293){.}
\put(251,294){.}
\put(251,294){.}
\put(252,295){.}
\put(253,296){.}
\put(254,297){.}
\put(254,298){.}
\put(255,299){.}
\put(256,300){.}
\put(257,301){.}
\put(258,302){.}
\put(259,303){.}
\put(260,304){.}
\put(261,305){.}
\put(262,306){.}
\put(262,307){.}
\put(263,308){.}
\put(264,309){.}
\put(265,310){.}
\put(266,311){.}
\put(267,312){.}
\put(87,43){.}
\put(90,46){.}
\put(92,49){.}
\put(94,52){.}
\put(97,55){.}
\put(99,58){.}
\put(103,63){.}
\put(106,66){.}
\put(108,69){.}
\put(111,72){.}
\put(113,75){.}
\put(115,78){.}
\put(118,81){.}
\put(120,84){.}
\put(123,87){.}
\put(125,90){.}
\put(127,93){.}
\put(130,96){.}
\put(132,99){.}
\put(136,104){.}
\put(139,107){.}
\put(141,110){.}
\put(143,113){.}
\put(146,116){.}
\put(148,119){.}
\put(152,124){.}
\put(155,127){.}
\put(157,130){.}
\put(159,133){.}
\put(162,136){.}
\put(164,139){.}
\put(168,144){.}
\put(171,147){.}
\put(174,150){.}
\put(176,153){.}
\put(179,156){.}
\put(181,159){.}
\put(185,164){.}
\put(188,167){.}
\put(190,170){.}
\put(193,173){.}
\put(195,176){.}
\put(198,179){.}
\put(201,183){.}
\put(204,186){.}
\put(206,189){.}
\put(208,192){.}
\put(211,195){.}
\put(213,198){.}
\put(217,203){.}
\put(220,206){.}
\put(222,209){.}
\put(225,212){.}
\put(227,215){.}
\put(230,218){.}
\put(233,222){.}
\put(236,225){.}
\put(238,228){.}
\put(241,231){.}
\put(243,234){.}
\put(246,237){.}
\put(250,241){.}
\put(252,244){.}
\put(255,247){.}
\put(258,250){.}
\put(260,253){.}
\put(263,256){.}
\put(266,260){.}
\put(269,263){.}
\put(271,266){.}
\put(274,269){.}
\put(276,272){.}
\put(279,275){.}
\put(283,279){.}
\put(285,282){.}
\put(288,285){.}
\put(291,288){.}
\put(293,291){.}
\put(296,294){.}
\put(299,297){.}
\put(301,300){.}
\put(304,303){.}
\put(307,306){.}
\put(309,309){.}
\put(312,312){.}
\put(81,41){.}
\put(82,43){.}
\put(87,49){.}
\put(88,50){.}
\put(89,51){.}
\put(90,53){.}
\put(95,59){.}
\put(96,60){.}
\put(97,61){.}
\put(98,63){.}
\put(99,64){.}
\put(104,70){.}
\put(105,71){.}
\put(106,72){.}
\put(107,74){.}
\put(112,80){.}
\put(113,81){.}
\put(114,82){.}
\put(115,84){.}
\put(120,90){.}
\put(121,91){.}
\put(122,92){.}
\put(123,94){.}
\put(128,100){.}
\put(129,101){.}
\put(130,102){.}
\put(131,104){.}
\put(136,110){.}
\put(137,111){.}
\put(138,112){.}
\put(139,114){.}
\put(144,120){.}
\put(145,121){.}
\put(146,122){.}
\put(147,124){.}
\put(152,130){.}
\put(153,131){.}
\put(154,132){.}
\put(155,134){.}
\put(160,140){.}
\put(161,141){.}
\put(162,142){.}
\put(163,143){.}
\put(164,145){.}
\put(169,150){.}
\put(170,152){.}
\put(171,153){.}
\put(172,154){.}
\put(177,160){.}
\put(178,161){.}
\put(179,162){.}
\put(180,163){.}
\put(185,169){.}
\put(186,171){.}
\put(187,172){.}
\put(188,173){.}
\put(193,179){.}
\put(194,180){.}
\put(195,181){.}
\put(196,183){.}
\put(201,189){.}
\put(202,190){.}
\put(203,191){.}
\put(204,193){.}
\put(209,199){.}
\put(210,200){.}
\put(211,201){.}
\put(212,202){.}
\put(217,208){.}
\put(218,210){.}
\put(219,211){.}
\put(220,212){.}
\put(225,218){.}
\put(226,219){.}
\put(227,220){.}
\put(228,221){.}
\put(229,222){.}
\put(234,228){.}
\put(235,229){.}
\put(236,230){.}
\put(237,231){.}
\put(242,237){.}
\put(243,238){.}
\put(244,239){.}
\put(245,240){.}
\put(250,246){.}
\put(251,248){.}
\put(252,249){.}
\put(253,250){.}
\put(258,256){.}
\put(259,257){.}
\put(260,258){.}
\put(261,259){.}
\put(266,265){.}
\put(267,267){.}
\put(268,268){.}
\put(269,269){.}
\put(274,275){.}
\put(275,276){.}
\put(276,277){.}
\put(277,278){.}
\put(282,284){.}
\put(283,285){.}
\put(284,286){.}
\put(285,287){.}
\put(290,293){.}
\put(291,294){.}
\put(292,295){.}
\put(293,296){.}
\put(298,302){.}
\put(299,303){.}
\put(300,304){.}
\put(301,305){.}
\put(306,311){.}
\put(307,312){.}
\put(149,20){\line(0,1){5}}
\put(149,10){13}
\put(263,20){\line(0,1){5}}
\put(263,10){14}
\put(20,81){\line(1,0){5}}
\put(10,81){23}
\put(20,141){\line(1,0){5}}
\put(10,141){24}
\put(20,200){\line(1,0){5}}
\put(10,200){25}
\put(20,260){\line(1,0){5}}
\put(10,260){26}
\end{picture}

{\center Figure 3. }

\pagebreak

\end{document}